\documentclass[
onecolumn, 
11pt, 
english,
tightenlines,
superscriptaddress, 
secnumarabic, 
nofootinbib, 
aps 
]{revtex4-2-jk} 

\usepackage[centering,hmargin=2.5cm,vmargin=2.5cm]{geometry}
\usepackage[colorlinks=true,citecolor=blue,linkcolor=blue]{hyperref}
\usepackage[normalem]{ulem}
\usepackage{amsmath,amssymb}
\usepackage{mathtools}
\usepackage{epsfig}
\usepackage{graphicx}               
\usepackage{url}
\usepackage{color}
\usepackage{multirow}
\usepackage{placeins}
\usepackage[dvipsnames]{xcolor}
\usepackage{epstopdf}
\usepackage{siunitx}
\usepackage{enumitem}
\usepackage{cleveref}
\usepackage{lipsum}
\usepackage{gensymb}
\usepackage{xspace}
\usepackage{titlesec}
 
\allowdisplaybreaks

\makeatletter
\renewcommand{\thesection}{\arabic{section}}
\renewcommand{\thesubsection}{\thesection.\arabic{subsection}}
\renewcommand{\p@subsection}{}
\makeatother

\titleformat*{\section}{\bfseries\large}
\titlelabel{\thetitle\quad}
\titleformat*{\paragraph}{\bfseries}
\titlespacing*{\paragraph}{0pt}{3.25ex plus 1ex minus .2ex}{1em}


\begin{document}

\pagenumbering{roman}

\title{\Large Neutrino Theory in the Precision Era \\[0.4cm]
       \large Summary Document of the CERN Neutrino Platform Pheno Week 2025 \\[0.2cm]
       \mdseries as Input to the \\[0.2cm]
       \bfseries European Strategy for Particle Physics 2025 Update}
\newcommand{\cern}{CERN, Geneva, Switzerland}
\newcommand{\IFIC}{Instituto de F\'{i}sica Corpuscular (CSIC-Universitat de Val\`{e}ncia), Parc Cient\'ific UV C/ Catedr\'atico Jos\'e Beltr\'an, 2 46980 Paterna (Valencia) - Spain}

\newcommand{\UB}{Departament de F\'isica Qu\`antica i Astrof\'isica and Institut de Ci\`encies del Cosmos, Universitat de Barcelona, Diagonal 647, E-08028 Barcelona, Spain}

\newcommand{\lisboa}{Centro de F\'isica Te\'orica de Part\'iculas -- CFTP and Dept de F\' \i sica, Instituto Superior T\'ecnico -- IST, Universidade de Lisboa, Av. Rovisco Pais, 1049-001 Lisboa, Portugal }

\newcommand{\UCL}{University College London (UCL), United Kingdom}

\newcommand{\FNAL}{Fermi National Accelerator Laboratory (FNAL), Batavia, IL 60510, USA}

\newcommand{\tum}{Physik-Department, Technische Universit{\"{a}}t, M{\"{u}}nchen, James-Franck-Stra{\ss}e, 85748 Garching, Germany}

\newcommand{\bilbao}{Department of Physics \& EHU Quantum Center, University of the Basque Country UPV/EHU, PO Box 644, 48080 Bilbao, Spain}

\newcommand{\mainz}{Johannes Gutenberg University Mainz, Germany}

\newcommand{\ihep}{Institute of High Energy Physics, Chinese Academy of Sciences, Beijing 100049, China}

\newcommand{\PRLAhmed}{Physical Research Laboratory, Ahmedabad, India}

\author{Asmaa~Abada}
\affiliation{P\^ole Th\'eorie, Laboratoire de Physique des 2 Infinis Irène Joliot Curie,
CNRS/IN2P3 et Universit\'e Paris Saclay, Orsay, France}

\author{Gabriela~Barenboim}
\affiliation{Departament de F\'{i}sica Te\`{o}rica, Universitat de Val\`{e}ncia, Burjassot 46100, Spain}
\affiliation{\IFIC}

\author{Toni~Bertólez-Martínez}
\affiliation{\UB}

\author{Sandipan~Bhattacherjee}
\affiliation{Indian Institute of Technology Madras, Chennai, India}
\affiliation{Birla Institute of Technology Mesra, India}

\author{Sara~Bolognesi}
\affiliation{IRFU, CEA, Universite' Paris-Saclay, F-91191 Gif-sur-Yvette, France}

\author{Patrick~D.~Bolton}
\affiliation{Jo\v zef Stefan Institute, Jamova 39, 1000 Ljubljana, Slovenia}

\author{Nilay~Bostan}
\affiliation{Department of Physics and Astronomy, University of Iowa, Iowa City, IA 52242, USA}
\affiliation{Proton Accelerator Facility, Turkish Energy Nuclear and Mineral Research Agency, Nuclear Energy Research Institute, 06980, Ankara, Türkiye}

\author{Gustavo~C.~Branco}
\affiliation{\lisboa}

\author{Sabya~Sachi~Chatterjee}
\affiliation{Karlsruhe Institute of Technology (KIT), Germany}

\author{Adriano~Cherchiglia}
\affiliation{Instituto de Fisica Gleb Wataghin, Universidade Estadual de Campinas, 13083-859, Campinas, SP, Brazil}

\author{Marco~Chianese}
\affiliation{Scuola Superiore Meridionale, Via Mezzocannone 4, 80138 Napoli, Italy}
\affiliation{INFN - Sezione di Napoli, Complesso Universitario Monte S. Angelo, 80126 Napoli, Italy}

\author{B.~A.~Couto~e~Silva}
\affiliation{Department of Physics and Helsinki Institute of Physics, University of Helsinki, P.O. Box 64, 00014 Helsinki, Finland.}
\affiliation{Departamento de F\'isica, UFMG, Belo Horizonte, MG 31270-901, Brazil.}

\author{Peter~B.~Denton}
\affiliation{High Energy Theory Group, Physics Department Brookhaven National Laboratory, Upton, NY 11973, USA}

\author{Stephen~Dolan}
\affiliation{\cern}

\author{Marco~Drewes}
\affiliation{Centre for Cosmology, Particle Physics, and Phenomenology, Université catholique de Louvain, Louvain-la-Neuve B-1348, Belgium}

\author{Ilham~El~Atmani}
\affiliation{Faculty of Sciences Ain-Chock, Hassan II University of Casablanca, Morocco}

\author{Miguel~Escudero}
\email{miguel.escudero@cern.ch}
\affiliation{\cern}

\author{Ivan~Esteban}
\affiliation{\bilbao}

\author{Manuel~Ettengruber}
\affiliation{Universite Paris-Saclay, CNRS, CEA, Institut de Physique Theorique, 91191, Gif-sur-Yvette, France}

\author{Enrique~Fern\'andez-Mart\'inez}
\affiliation{Instituto de F\'{\i}sica Te\'orica UAM/CSIC, C/ Nicol \'as Cabrera 13-15, Universidad Aut\'onoma de Madrid, Cantoblanco, 28049 Madrid, Spain}

\author{Julien~Froustey}
\affiliation{Department of Physics, University of California Berkeley, Berkeley, CA 94720, USA}
\affiliation{Department of Physics, University of California San Diego, La Jolla, CA 92093, USA}

\author{Raj~Gandhi}
\affiliation{Harish Chandra Research Institute (HRI)}
\affiliation{Jhusi, Chhatnag Road, Prayagraj, UP 211019, India}

\author{Julia~Gehrlein}
\affiliation{Physics Department, Colorado State University, Fort Collins, CO 80523, USA}

\author{Srubabati~Goswami}
\affiliation{\PRLAhmed}

\author{Andr\'e~de~Gouv\^ea}
\affiliation{Department of Physics and Astronomy, Northwestern University, USA}

\author{Alessandro~Granelli}
\affiliation{Dipartimento di Fisica e Astronomia, Universit\`{a} di Bologna/INFN, via Irnerio 46, 40126, Italy}

\author{Rasmi~Hajjar}
\affiliation{\IFIC}

\author{Pilar~Hernández}
\affiliation{\IFIC}

\author{Gonzalo~Herrera}
\affiliation{Center for Neutrino Physics, Department of Physics, Virginia Tech, Blacksburg, VA 24061, USA}

\author{Matheus~Hostert}
\affiliation{Department of Physics \& Laboratory for Particle Physics and Cosmology, Harvard University, Cambridge, MA 02138, USA}

\author{Alejandro~Ibarra}
\affiliation{\tum}

\author{Yu~Seon~Jeong}
\affiliation{Department of Physics, Institute of Basic Science, Sungkyunkwan University, Suwon 16419, Korea}

\author{Filipe~R.~Joaquim}
\affiliation{\lisboa}

\author{Monireh Kabirnezhad} 
\affiliation{Imperial College London, Department of Physics, London SW7 2BZ, United Kingdom}
 
\author{Kevin~J.~Kelly}
\affiliation{Department of Physics and Astronomy, Mitchell Institute for Fundamental Physics and Astronomy, Texas A\&M University, College Station, TX 77843, USA}

\author{Pyungwon~Ko}
\affiliation{School of Physics, Korea Institute for Advanced Study, Seoul 02455, Republic of Korea}

\author{Joachim~Kopp}
\email{jkopp@cern.ch}
\affiliation{\mainz}
\affiliation{\cern}

\author{Zoha~Laraib}
\affiliation{University of Tennessee, Knoxville}

\author{Shirley~Li}
\affiliation{Department of Physics and Astronomy, University of California, Irvine, CA 92697, USA}

\author{Chayan~Majumdar}
\affiliation{\UCL}

\author{Xabier~Marcano}
\affiliation{\bilbao}

\author{Danny~Marfatia}
\affiliation{Department of Physics and Astronomy, University of Hawaii, Honolulu, HI 96822, USA}

\author{Hyun~Min~Lee}
\affiliation{Department of Physics, Chung-Ang University, Seoul 06974, Republic of Korea}

\author{Manimala~Mitra}
\affiliation{Institute of Physics, Bhubaneswar 751005, India}

\author{Rukmani~Mohanta}
\affiliation{School of Physics, University of Hyderabad, Hyderabad - 500046, India}

\author{Biswarup~Mukhopadhyaya}
\affiliation{Indian Institute of Science Education and Research Kolkata, India}

\author{Maksym~Ovchynnikov}
\email{maksym.ovchynnikov@cern.ch}
\affiliation{\cern}

\author{Supriya Pan}
\affiliation{\PRLAhmed}

\author{Ornella~Palamara}
\affiliation{\FNAL}

\author{Stephen~J.~Parke}
\affiliation{\FNAL}

\author{George~A.~Parker}
\affiliation{\mainz}

\author{Silvia~Pascoli}
\affiliation{Dipartimento di Fisica e Astronomia, Universit\`{a} di Bologna/INFN, via Irnerio 46, 40126, Italy}
\affiliation{INFN, Sezione di Bologna, viale Berti Pichat 6/2, 40127 Bologna, Italy}

\author{Joselen~Pena~Quintero}
\affiliation{Department of Physics, Universidad de Antofagasta
Antofagasta 1240000, Chile}

\author{Jo\~ao~Paulo~Pinheiro}
\affiliation{\UB}

\author{Federica~Pompa}
\affiliation{SUBATECH - CNRS/IN2P3}

\author{Yago~Porto}
\affiliation{Centro de Ciencias Naturais e Humanas, Universidade Federal do ABC, 09210-170, Santo Andre, SP, Brazil}
\affiliation{Instituto de Fisica Gleb Wataghin, Universidade Estadual de Campinas, 13083-859, Campinas, SP, Brazil}

\author{Suraj~Prakash}
\affiliation{\IFIC}

\author{M.~N.~Rebelo}
\affiliation{\lisboa}

\author{Albert~de~Roeck}
\affiliation{\cern}

\author{Juan~Rojo}
\affiliation{Department of Physics and Astronomy, VU Amsterdam, NL-1081 HV Amsterdam, The Netherlands}

\author{Valentina~De~Romeri}
\affiliation{\IFIC}

\author{Salvador~Rosauro-Alcaraz}
\affiliation{INFN, Sezione di Bologna, viale Berti Pichat 6/2, 40127 Bologna, Italy}

\author{Purushottam~Sahu}
\affiliation{Indian institute of Technology Bombay,400076 Mumbai, Maharashtra India}

\author{Ina~Sarcevic}
\affiliation{University of Arizona, USA}

\author{Ninetta~Saviano}
\affiliation{INFN - Sezione di Napoli, Complesso Universitario Monte S. Angelo, 80126 Napoli, Italy}
\affiliation{Scuola Superiore Meridionale, Via Mezzocannone 4, 80138 Napoli, Italy}

\author{Michael~A.~Schmidt}
\affiliation{School of Physics, The University of New South Wales, Sydney NSW 2052, Australia}

\author{Ian M.~Shoemaker}
\affiliation{Center for Neutrino Physics, Department of Physics, Virginia Tech, Blacksburg, Virginia 24601, USA}

\author{Alka~Singh}
\affiliation{Department of Physics, Aligarh Muslim University, Aligarh, India - 202002}

\author{Zahra~Tabrizi}
\email{z\_tabrizi@pitt.edu}
\affiliation{Pittsburgh Particle Physics Astrophysics and Cosmology Center, University of Pittsburgh, USA}
\affiliation{\cern}

\author{S.~Uma Sankar}
\affiliation{Department of Physics,
             Indian Institute of Technology Bombay,
             Mumbai 400076, India.}

\author{Salvador Urrea}
\affiliation{IJCLab, Pôle Théorie (Bat. 210), CNRS/IN2P3, 91405 Orsay, France}

\author{Zoya~Vallari}
\affiliation{Department of Physics, Ohio State University, Columbus, OH 43210, USA}

\author{Biao~Wang}
\affiliation{Department of Physics and Astronomy, University of Iowa, Iowa City, IA 52242, USA}

\author{Xin~Wang}
\affiliation{School of Physics and Astronomy, University of Southampton, Southampton, United Kingdom}

\author{Zhi-zhong Xing}
\affiliation{\ihep}

\author{Farhana~Zaidi}
\affiliation{Department of Physics, Aligarh Muslim University, Aligarh, India - 202002}

\author{Di~Zhang}
\affiliation{\tum}

\author{Zhong~Zhang}
\affiliation{\UCL}

\author{Shun~Zhou}
\affiliation{\ihep}

\collaboration{see appendix for authors' affiliations}
\noaffiliation

\date{\today}

\begin{abstract}
This document summarises discussions on future directions in theoretical neutrino physics, which are the outcome of a neutrino theory workshop held at CERN in February 2025. The starting point is the realisation that neutrino physics offers unique opportunities to address some of the most fundamental questions in physics. This motivates a vigorous experimental programme which the theory community fully supports. \textbf{A strong effort in theoretical neutrino physics is paramount to optimally take advantage of upcoming neutrino experiments and to explore the synergies with other areas of particle, astroparticle, and nuclear physics, as well as cosmology.} Progress on the theory side has the potential to significantly boost the physics reach of experiments, as well as go well beyond their original scope. Strong collaboration between theory and experiment is essential in the precision era. To foster such collaboration, \textbf{we propose to establish a CERN Neutrino Physics Centre.} Taking inspiration from the highly successful LHC Physics Center at Fermilab, the CERN Neutrino Physics Centre would be the European hub of the neutrino community, covering experimental and theoretical activities.
\end{abstract}

\maketitlenoaffil

\clearpage

\pagenumbering{arabic}
\setcounter{page}{1}

\section{Executive Summary}

From February 17--21 2025, CERN hosted the \href{https://indico.cern.ch/event/1454726/}{Neutrino Platform Pheno Week 2025}, the 5th workshop in a series focused on new developments in neutrino theory and phenomenology. Part of the 2025 edition was a discussion about the future strategy of the field in Europe and worldwide. This document presents the outcome of this discussion.

It can be subsumed into three key messages:
\begin{enumerate}
    \item Neutrino physics offers unique strategic opportunities. Neutrinos allow us to address some of the most fundamental questions in physics, including the origin of the matter--antimatter asymmetry of the Universe, the mechanism generating neutrino masses, the origin of the flavour structure of elementary particles, the nature of the highest-energy cosmic ray sources, and the dynamics of supernovae. In addition, neutrinos can test a broad range of extensions of the Standard Model. \textbf{We strongly support a vigorous programme addressing these questions, both experimentally and theoretically.} 
    
    In a broader context, capitalizing on the data from upcoming large-scale neutrino experiments will allow the European community to benefit from the investments it has made into the hardware of these experiments and to maintain a strong particle physics programme at the forefront of the field during a time before a new collider becomes available.

    \item \textbf{A strong effort in theoretical neutrino physics is paramount to optimally take advantage of upcoming neutrino experiments and to explore the synergies with other areas of particle, astroparticle and nuclear physics, as well as cosmology.} Progress on the theory side has the potential to significantly boost the physics reach of experiments, as well as go well beyond their original scope.

    \item Strong collaboration between theory and experiment is essential in the precision era. To foster such collaboration, \textbf{we propose to establish a CERN Neutrino Physics Centre.} Taking inspiration from the highly successful LHC Physics Center at Fermilab, the CERN Neutrino Physics Centre would be the European hub of the neutrino community, covering experimental and theoretical activities.
\end{enumerate}

\section{Introduction}

Neutrinos, once the elusive oddballs of the Standard Model, have matured into precision tools that may offer answers to some of the most fundamental questions in particle physics, astrophysics, and cosmology. Notably, neutrinos will allow us to significantly advance our theoretical understanding of nature by addressing the following pressing questions:
\begin{itemize}
    \item \emph{Is the CP symmetry violated in the lepton sector?} Given the role that symmetries play in the Standard Model, the violation of the CP symmetry is of far-reaching theoretical importance, and the lepton sector could provide a new key insight on its origin. Moreover, given the smallness of CP violation in the quark sector, new sources of CP violation are needed to explain the matter--antimatter asymmetry of the Universe. Leptonic CP violation could play this role.
    
    \item \emph{What is the origin of flavour?} The observation of three fermion generations, the peculiar structure of fermion masses, and the smallness of neutrino masses call for an explanation. The fact that quark and lepton mixings differ substantially offers the opportunity to address this fundamental problem from different and complementary angles. Given that the physics responsible for these structures is likely beyond the reach of high-energy colliders, precision measurements, combined with the development of new theoretical mechanisms and models, are the best handle we have for making progress in this field.
    
    \item \emph{Are neutrinos their own antiparticles?} The Standard Model does not tell us how neutrinos obtain their masses. In fact, they are the only known particles that could have a Majorana mass term, which would make them their own antiparticles. On top of this, the Majorana mass of neutrinos is the key to understand whether the lepton number symmetry is violated in nature. Majorana masses arise naturally in the seesaw mechanism that explains the smallness of neutrino masses and simultaneously offers an elegant explanation of the particle--antiparticle asymmetry of the Universe via leptogenesis.

    \item \emph{What is the inner structure of nucleons and atomic nuclei?} Neutrino--nucleus interactions probe properties of strongly interacting systems that are inaccessible in lepton scattering or in high-energy collisions.

    \item \emph{What lies beyond the Standard Model?} While the shortcomings of the Standard Model are evident, the best way of going beyond is not. Neutrinos play an important role in the global search for physics beyond the Standard Model, offering opportunities complementary to those available at high-energy colliders, quark flavour physics experiments, and fixed-target experiments.

    \item \emph{What is the dynamics of supernovae?} In spite of tremendous progress in simulating the cataclysmic explosions in which massive stars end their lives, supernovae are still an enigma. The observation of neutrinos from the next galactic supernova will be a game-changer in this field, but relating the neutrino signals detected at Earth to the dynamics of the supernova core requires a theoretical understanding of how neutrinos propagate out of the exploding star. While great progress has been made in the previous decades, this understanding, especially insofar as it concerns the neutrino flavour evolution, is lacking so far. 

    \item \emph{What is the origin of the highest-energy cosmic rays?} With the combination of advanced neutrino telescopes, gravitational wave observatories, ultra high energy cosmic ray telescopes, and sophisticated theoretical modelling of astrophysical neutrino sources we may be on the verge of answering this century-old question.

    \item \emph{How did the Universe evolve?} Neutrinos are the most abundant fermions in the Universe, and they influence its history in important ways. Their role is becoming even more crucial in the context of ongoing galaxy surveys mapping the large-scale structure whose theoretical interpretation crucially depends on neutrino properties. In particular, current cosmological observations can be used to place world-leading limits on the absolute neutrino mass scale. 

\end{itemize}
The experimental tools to address this list of questions -- which is far from complete -- are the new flagship neutrino oscillation experiments currently under construction in North America (DUNE) and Asia (HyperKamiokande, JUNO), complemented by a multifaceted programme of small and medium-scale experiments, many of them in Europe.

\textbf{We strongly endorse the planned experimental programme as we believe that it offers the optimal way forward towards answering the above questions, and many others.}

\textbf{An equally vigorous effort is required on the theory side. Firstly, strong theoretical support is needed to support and exploit experiments to reach their physics goals. Secondly, theoretical developments are needed to realise the synergies between different classes of neutrino experiments, and between neutrino physics and other areas of particle, astroparticle and nuclear physics, as well as cosmology. Finally, progress on the theory side has the potential to greatly enhance the physics reach of experiments, allowing them to go significantly beyond their original scope.}

In the following sections, we will outline the focus areas of theoretical neutrino physics in the coming years. We will conclude in \cref{sec:cnpc} by outlining the idea of a CERN Neutrino Physics Centre as a support infrastructure enabling, bundling, and enhancing these efforts.

\section{Leptonic Mixing and CP Violation}

The discovery of neutrino oscillations at the end of the twentieth century has revolutionised our understanding of neutrinos and has stimulated a comprehensive experimental programme to exploit this new phenomenon in order to reveal the flavour structures that lie hidden in the leptonic sector and to open a new window to physics beyond the Stadard Model. 

Ongoing and near-future precision measurements of neutrino oscillations will serve as a powerful probe of the completeness of the so-called three-massive-neutrinos paradigm, the hypothesis that there are exactly three massive neutrinos and that their interactions are those prescribed by the Standard Model. Deviations of neutrino oscillation patterns relative to those predicted by the three-massive-neutrinos paradigm would represent a clear signal of new physics in the neutrino sector. Simple hypotheses include the existence of light, Standard Model singlet fermions, new neutrino interactions, or unexpectedly enhanced neutrino properties like electromagnetic dipole moments or decay widths (see also \cref{sec:bsm} for a discussion of new physics in the neutrino sector). Key to uncovering such deviations is the concurrent precision measurement of different oscillation channels, e.g.,\ $\nu_{\mu}\to \nu_e$ and $\nu_{\mu}\to \nu_{\tau}$, and using different experimental setups -- different neutrino sources, different baselines, neutrinos versus antineutrinos, etc.  

Independently of the potential to unravel BSM phenomena in neutrino oscillations, precision measurements of neutrino oscillations are the only means to measure mixing in the leptonic sector. This, in turn, may prove to be invaluable in addressing the elusive flavour puzzle: why are there three generations? Why are the fermion masses so different from one another? why is quark mixing ``small'' and hierarchical, while lepton mixing appears to be ``large''? Over the years, the community has proposed a diverse and extensive list of flavour models. Many of these make concrete predictions about relations among the different elements of the leptonic and quark mixing matrices and masses. Tests of these models require precision measurements of the neutrino oscillation parameters, both the elements of the leptonic mixing matrix and the neutrino mass-squared differences. 

Of central importance are decisive measurements of CP violation in the lepton sector. To date, all observed CP-violating phenomena are governed by the CP-odd phase in the quark mixing matrix. The observation of CP-violation in neutrino oscillations would establish, for the first time, the existence of additional sources of CP violation in fundamental physics. It is well known that the CP-violating physics in the quark sector is insufficient for baryogenesis and that new, independent sources are required. In this sense, the discovery of CP violation in the lepton sector may contain the key to the matter--antimatter asymmetry of the Universe.  

The current and next-generation of neutrino oscillation experiments is expected to determine the neutrino mass ordering: normal ordering versus inverted ordering. The neutrino mass ordering plays a central role, together with the measurement of the mass-squared differences and the mixing parameters, in translating the results of other probes of neutrino mass -- precision measurements of $\beta$-decay, searches for neutrinoless double-beta decay, and measurements of the large-scale structure of the Universe -- into direct information on the values of the individual neutrino masses, addressing the Majorana versus Dirac question, and testing the consistency among this diverse set of observables. The neutrino mass ordering may also be intimately related to the dynamics behind nonzero neutrino masses. If, for example, the neutrino mass ordering is inverted, we will learn that the two largest neutrino masses are quasi-degenerate (at the few percent level). This quasi-degeneracy is a robust prediction of a subset of neutrino-mass and lepton-flavour models. 

In the coming years, the neutrino theory community will utilise precision measurements of neutrino oscillations to
\begin{itemize}
    \item Assess the compatibility of new measurements with the predictions of the three-massive-neutrino paradigm.

    \item Test models of flavour against the data.

    \item Develop advanced analysis strategies in close collaboration with the experimental community. This includes new ways of analysing the combined data from near and far detectors, different oscillation channels, and different interaction channels. It moreover includes combined fits of multiple experiments, which are essential for finding possible anomalies.
   
    \item Use precision oscillation measurements to search for new physics (see also \cref{sec:bsm}).
\end{itemize}

\section{Neutrino--Nucleus Interactions}

Accelerator-based long-baseline neutrino oscillation experiments -- the flagships of the field -- will soon reach the point where their measurements become limited by systematic uncertainties. By far the most important uncertainties are those associated with neutrino production rates and neutrino--nucleus scattering cross-sections. Advanced experimental techniques (such as near detectors capable of measuring on-axis and off-axis rates) can partially mitigate these uncertainties, but ultimately our limited theoretical understanding of how neutrinos interact with nuclei remains a crucial bottleneck.

Accelerator-based neutrino experiments typically operate in the energy range from hundreds of MeV to tens of GeV, implying that neutrino interactions need to be understood across several qualitatively distinct regimes. At the highest energies, above several GeV, the scattering is predominantly deep-inelastic (DIS), admitting a description in terms of parton distribution functions (PDFs). At energies around 1--3\,GeV, neutrino scattering can excite nucleon resonances like the $\Delta(1232)$, while at even lower energies quasi-elastic scattering dominates. An important additional contribution in the GeV range comes from so-called two-particle--two-hole (2p2h) processes in which a neutrino effectively scatters on a two-nucleon cluster. At all energies (even in the DIS regime due to the EMC effect), nuclear in-medium effects are crucial: they control the initial state of the struck nucleon, and they affect the outgoing particles as they propagate out of the nucleus. It is particularly important to have the description of exclusive states at all energies as well as accurately understanding the differences between the interactions of $\nu_e$, $\bar{\nu}_e$, $\nu_\mu$, and $\bar{\nu}_\mu$ neutrinos.

Going beyond long-baseline experiments, additional challenges exist: neutrino telescopes sensitive at energies from TeV--EeV probe kinematic regions in which the PDFs have never been measured. Lower-energy searches at nuclear reactors probe specific nuclear reactions and/or coherent elastic neutrino--nucleus scattering (CE$\nu$NS).

Future priorities for the neutrino theory community in the field of neutrino--nucleus interactions include
\begin{itemize}    
    \item Modelling medium-sized nuclei, in particular Ar-40, the target material in DUNE. Most existing nuclear models work best for light nuclei (C-12, O-16) and are difficult to extend to higher nuclear masses.

    \item Improving our understanding of the role of multi-nucleon effects on neutrino--nucleus scattering. This includes short-range correlations, two-particle--two-hole (2p2h) processes, as well as higher-order processes (e.g.\ 3p3h).

    \item Understanding the transition region between perturbative and non-perturbative regimes that govern neutrino scattering, bridging partonic and hadronic degrees of freedom, and testing quark–hadron duality in weak interactions. Lattice QCD techniques and higher-order QCD calculations are of great importance in this context. It will be crucial to have reliable predictions not only for inclusive cross sections, but also for exclusive processes.

    \item Assessment of theoretical uncertainties on neutrino cross sections. The theory work devoted to the improvement of these cross sections should also reflect uncertainties.

    \item Improved predictions of neutrino beam fluxes and spectra. In particular, it is critical to improve the modelling of hadroproduction to control the flux for neutrinos at long baseline experiments as well as the atmospheric neutrino flux. Data from experiments like NA61/SHINE and ENuBET+NuTAG will play a critical role in this effort.

    \item Improved event generators. Many recent advances in the modelling of neutrino--nucleus interactions have not been implemented in state-of-the-art event generators yet, a clear shortcoming of these generators. This includes models of 2p2h scattering, modern descriptions of the nuclear initial state, and higher-order effects for neutrino--nucleus scattering at and above LHC energies. In addition, improved modelling of final-state interactions is highly desirable. On the technical side, structured and continual development of the generators' computational backbone will be even more crucial in the future than it is now.
    
    \item DIS cross-sections and PDFs at energies from tens of GeV (SHiP) over TeV (FASER, SND@LHC, Forward Physics Facility) to EeV (neutrino telescopes). In this context, neutrinos can in particular serve as unique probes of QCD matter, constraining the structure of the proton in novel kinematic regimes.

    \item Close theory--experiment collaboration. Through continued exchange with the experimental community, theoretical studies can optimally address specific experimental issues such as the energy carried away by neutrons in a neutrino--nucleus interaction, the multiplicity of pions in the final state, or the energy dependence of the cross sections. In addition, a strong theoretical effort is needed to optimally utilize auxiliary measurements (DUNE-PRISM, NA61/SHINE, EnuBET+NuTAG, nuSTORM, etc.) to improve neutrino interactions models.
\end{itemize}

\section{Neutrino Nature and Absolute Neutrino Masses}

While neutrino oscillations prove that neutrinos have mass, the absolute scale of these masses as well as the mechanism by which they are generated, remain a mystery. It is expected that cosmology will determine the absolute neutrino mass scale in the coming years, while direct laboratory measurements (currently led by KATRIN, but with key projects such as Project~8, KATRIN++ or ECHO/HOLMES in the future) still have technological challenges to overcome before achieving the same. When they do, however, their results will be more robust with regard to potential deviations from standard cosmology. Theoretical developments are crucial in the quest for neutrino mass measurements, especially in the context of cosmology. For instance, the power of large-scale structure surveys stems to a large extent from the development of cosmological perturbation theory, which allows to relate observations to the underlying theoretical model in an efficient way.

The second big question in the context of neutrino masses is the theoretical origin of the corresponding mass terms in the Lagrangian. Neutrinos are the only fermions in the Standard Model that could be their own anti-particles (which would make them Majorana fermions), and in fact this is what is predicted in most theoretical models aiming to explain the smallness of neutrino masses. Establishing that neutrinos are indeed Majorana fermions would not only constitute the discovery of an entirely new type of fundamental particle, but would simultaneously constitute a measurement of the absolute neutrino mass scale. In addition, it would prove that lepton number -- a fundamental, albeit accidental, symmetry in the Standard Model -- is broken. The only known way of achieving this in practice is by observing neutrinoless double beta ($0\nu2\beta$) decay, one of the rarest processes in the Universe. Besides establishing the Majorana nature of neutrinos, $0\nu2\beta$ decay is also the only realistic experimental handle we have on the Majorana phases in the PMNS matrix.  It comes, however, with formidable experimental challenges, especially if neutrino masses turn out to follow a normal hierarchy, in which case the rate of $0\nu2\beta$ decay is significantly smaller than in the inverted hierarchy case. In addition, it is hampered by large $\mathcal{O}(1)$ theoretical uncertainties on the nuclear matrix elements which govern the decay rate. Reducing these uncertainties is crucial not only for interpreting experimental results, but also for the planning of future detectors as the target mass required to achieve a certain sensitivity to neutrino properties depends strongly on the values of the relevant nuclear matrix elements.

In view of the above, future priorities in the field include
\begin{itemize}
    \item Optimally exploit the data from current and upcoming cosmological surveys. This requires in particular a deep understanding of how neutrinos affect the Large Scale Structure (LSS) distribution of galaxies and quasars in the Universe. Cosmological simulations with neutrinos are notoriously difficult and a first principles implementation of neutrinos within the Effective Field Theory (EFT) of LSS is still lacking. 
    
    \item Improved matrix elements for $0\nu2\beta$. In recent years, new methods have been developed to model the nuclei of interest, leading to significant new insights such as the existence of a previously overlooked short-distance contribution to the matrix elements. In addition, auxiliary measurements have been proposed to constrain nuclear models. Further developing these concepts is of very high importance to the field.

\end{itemize}

\section{Neutrino Astrophysics}

Immediately after the discovery of neutrinos, their extraordinary utility as astrophysical probes has been realized, first in the context of solar neutrinos. By now, neutrinos have allowed us to investigate the interior of the Sun, the dynamics of supernova explosions, and the acceleration of ultra-high energy cosmic rays in active galactic nuclei, blazars, and our own Milky Way. All observations of astrophysical neutrinos have spurred dramatic progress in theory and phenomenology, including on theoretical modelling of the sources, neutrino flavour evolution, and often game-changing searches for physics beyond the Standard Model. While it is clear that observations of astrophysical neutrinos will invigorate theoretical research in the same way also in the future, theoretical work is also becoming increasingly important to enable these observations in the first place.

In the coming years, the priorities of the theoretical neutrino astrophysics community will include
\begin{itemize}
    \item Advanced modelling of ultra-high-energy astrophysical neutrino point sources such as active galactic nuclei to constrain the properties of non-thermal particle acceleration in environments which cannot be probed by any other messenger.

    \item Improved understanding of the interactions of ultra-high energy neutrinos with nuclei inside or around a detector, combined with better modelling of backgrounds, in particular the one due to prompt atmospheric neutrinos, where LHC neutrino measurements will play a crucial role.

    \item The flavour evolution of supernova neutrinos. Neutrinos propagating out of an exploding star strongly influence each other thanks to their enormous density $> 10^{30}\,\text{cm}^{-3}$. Because of this, their evolution is highly non-linear and strongly influenced by inhomogeneities and flavour instabilities. Since neutrinos affect the explosion itself, this makes it impossible to reliably predict the expected signal from the next galactic supernova, severely hindering our chances at using this unique multi-messenger opportunity. Progress in this field is expected with advanced theoretical and computational methods, including potentially quantum computing. In the longer term, it is highly desirable to include robust modelling of neutrino flavour evolution directly in hydrodynamic simulations of the explosion.

    \item Neutrinos from compact stars. Neutrinos are emerging as a novel tool to study neutron stars and black holes. For instance, the fact that some neutron stars appear to cool anomalously fast calls for better understanding of neutrino emission -- the main cooling channel -- in these stars. Moreover, it has been suggested that neutrinos can probe extreme events such as common-envelope phases of binary evolution.

    \item New physics searches with astrophysical neutrinos, including for instance searches for neutrino--dark matter interactions, neutrino self-interactions, and even CPT and Lorentz symmetry violation.
    
\end{itemize}

\section{Neutrino Cosmology}

Neutrinos were produced copiously in the early Universe after the Big Bang, and their presence has profound cosmological implications. In particular, since these neutrinos have non-zero mass, they gravitate and represent a non-negligible contribution ($\gtrsim 0.14\%$) to the energy density of the Universe today. Current galaxy surveys and observations of the Cosmic Microwave Background (CMB) have now the sensitivity to measure this tiny energy density and determine the absolute neutrino mass scale as a result. Furthermore, high precision CMB observations will provide unprecedented measurements of the effective number of relativistic neutrino species in the early Universe ($N_{\rm eff}$) leading to either support for the Standard Model of cosmology or to a clear hint beyond. From the theory side, neutrinos may be the key behind two puzzles in cosmology: the nature of dark matter, and the origin of the matter--antimatter asymmetry in the Universe. In particular, the leptogenesis scenario elegantly uses the heavy sterile neutrino partners to generate an imbalance between matter and antimatter in the early Universe.

In view of the exciting opportunities offered by upcoming cosmological observations, key priorities in the field are:
\begin{itemize}
    \item Optimise the exploitation of upcoming surveys to measure neutrino masses. A full theoretical understanding of the impact of massive neutrinos in the distributions of galaxies in the Universe is still lacking. This description is non-trivial due to their high velocities as compared with cold dark matter. This in turn make numerical simulations including neutrinos very challenging, and also imply that the use of EFT of LSS in the context of neutrinos is not straightforwardly applied either. Furthermore, cosmological neutrino mass bounds depend on the cosmological model and it will be important to understand their robustness upon modifications of the cosmological model.

    \item Understand the ``Hubble tension'' and its connections to neutrino physics. Local (low-redshift) measurements of the Hubble constant seem to indicate that Universe is expanding faster that predicted based on observations from the early Universe. This so-called Hubble tension is closely connected to neutrino physics: given CMB observations, a larger Hubble constant points towards smaller neutrino masses.  In addition, the large significance of the tension ($> 5\sigma$) questions the validity of the Standard Model of cosmology and calls into doubt its applicability for inferring fundamental parameters such as the neutrino mass. Theory work trying to resolve this tension is of critical importance for progress in cosmology.

    \item Connections between neutrinos and dark matter. This includes the possibility that heavy sterile neutrinos \emph{are} the dark matter (a scenario that can be tightly constrained by looking for signals of dark matter decay), as well as neutrino--dark matter interactions.

    \item Origin of matter via Leptogenesis. Decays of heavy sterile neutrinos, as well as oscillations involving sterile neutrinos, provide elegant mechanisms to explain why the Universe contains more matter than anti-matter. While the simplest scenarios of this type operate at very high temperatures in the early Universe, others feature light states that could be accessible to laboratory experiments such as the LHC, SHIP and FCC. It is therefore important to understand the testability of these scenarios, particularly for models with sterile neutrinos with $M < 100\,{\rm GeV}$. For those scenarios operating at  higher energies, it would be interesting to also explore other signals such as gravitational waves.

    \item The cosmic neutrino background (C$\nu$B). Directly observing the population of cosmic relic neutrinos is considered the holy grail of neutrino cosmology, but the low energy of these neutrinos presents a unique challenge. It is therefore crucial to explore new avenues for C$\nu$B detection both experimentally and theoretically, as to thoroughly examine the potential of the ones already under consideration. 
\end{itemize}

\section{Physics Beyond the Standard Model}
\label{sec:bsm}

The Standard Model of particle physics is incomplete and leaves many key questions unanswered: What is the origin of neutrino masses? What is the nature of dark matter? Why is the Universe made of only matter? These questions motivate the presence of new particles and fields beyond those currently known, and in many scenarios these model extensions are related to neutrinos, or neutrinos offer a unique way of testing them. This motivates comprehensive searches for physics beyond the Standard Model in the neutrino sector, connecting the vigorous experimental programmes in neutrino physics with direct or indirect signals in collider experiments, dark matter searches, astrophysical observations, and cosmology. Given the exciting experimental opportunities on these various fronts, priorities of the field include:
%
%
\begin{itemize}   
    \item Exploit the BSM potential of long baseline neutrino oscillation experiments. This includes interpreting data from both the far and near detectors, predicting signals, and identifying new opportunities.  Possible BSM targets include light or heavy sterile neutrinos, light dark matter particles, axion-like particles, and novel neutrino interactions either in an effective field theory framework or in the context of light mediators.
    
    \item Improved neutrino interaction cross-sections. Potential BSM signal in experiments such as DUNE or HyperKamiokande often affect the near and far detectors differently. The cancellation of systematic uncertainties between the two detectors is then spoiled, reducing the sensitivity to both standard oscillations and new physics. In this context, improving theoretical predictions of neutrino--nucleus interaction cross sections is even more imperative than in the absence of new physics. This includes incorporating the manifold ways in which neutrino interactions may be modified by physics beyond the Standard Model.
    
    \item Leverage the complementarity between different neutrino experiments, including long-baseline and short-baseline accelerator-based experiments, searches for neutrinoless double beta decay, collider neutrino experiments, coherent elastic neutrino--nucleus scattering, direct neutrino mass measurements, neutrino astrophysics, and neutrino cosmology.
        
    \item Leverage the complementarity with other BSM searches. While neutrinos offer a unique window to many possible extensions of the Standard Model, a full picture can only emerge when neutrino measurements are combined with data from ongoing and future collider and beam dump experiments, quark and charged lepton flavour physics measurements, as well as experiments in astroparticle physics and cosmology. In particular, models explaining dark matter or addressing the flavour puzzle and neutrino masses predict signals not only in neutrino experiments, but they may also feature new states close and below the electroweak scale, accessible to the LHC, SHiP, or a future collider.  Interpreting these vastly different data sets requires powerful theoretical tools such as effective field theories.

    \item Event generators for BSM physics. As the BSM programme at neutrino experiments expands, it will be essential to have powerful Monte Carlo event generators capable of simulating signals beyond the SM.

    \item Foster close collaboration between theory and experiment. Searches for physics beyond the Standard Model depend crucially on such collaboration to identify targets that are theoretically well-motivated and interesting, while at the same time experimentally reachable. Similarly, the interpretation of data in the context of BSM models relies on input from both experiment and theory, and on efficient ways of sharing data in such a way that it can be recast and repurposed in a meaningful way.    
\end{itemize}

\section{The CERN Neutrino Physics Centre}
\label{sec:cnpc}

As we have outlined in the preceding sections, there are key open questions in neutrino physics that can be answered within the next years and which will need crucial theoretical input. In this context, strong collaboration between neutrino theory and experiment is indispensable in the precision neutrino era. \emph{To foster such collaboration between theory and experiment, we propose to establish a CERN Neutrino Physics Centre. Taking inspiration from the highly successful LHC Physics Center at Fermilab, the CERN Neutrino Physics Centre would be the European hub of the neutrino community, covering experimental and theoretical activities.} 

The key components of the centre would be
\begin{itemize}
    \item Infrastructure for organising workshops. These will range from conferences covering a broad range of topics related to neutrinos, over collaboration meetings, to small focused workshops and meetings of individual working groups. The larger meetings will allow the community to exchange ideas and results in an interdisciplinary fashion, while smaller ones will provide an environment for solving specific open problems.

    \item Training early career researchers. The drivers of the field are early career researchers, and their education is pivotal to finding answers to the open questions above. The centre will organise schools and training programmes to equip the current and next generation of researchers with state-of-the art tools.
    
    \item Support for short-term and long-term visits to CERN. Neutrino physics is cross-disciplinary and progress often requires strong collaboration of experts from different fields. As these experts are often geographically far apart, a strong visitor programme will enable key subsets of the community to meet at CERN. The centre would thereby foster collaboration and generate the scientific environment where major breakthroughs can occur.

    \item A strong theory component. In a field where many crucial measurements are limited by theoretical uncertainties, such as those associated with neutrino--nucleus interaction cross sections or with neutrino flavour evolution in supernovae, modest investments into theoretical research can have a transformative impact. While Europe has always had a very strong neutrino theory community, this community is somewhat dispersed and would therefore greatly benefit from a central hub at CERN. We envision the CERN Neutrino Physics Centre as the place where researchers can develop new ideas, collaborate across disciplines, and ultimately make key contributions that will enable ground-breaking discoveries. 

    \item Dedicated computing resources. Solving open theoretical problems such as the modelling of neutrino--nucleus interactions or neutrino oscillations in supernovae require intensive numerical computations. Similarly, also experimental computing needs are increasing rapidly. In this context, the CERN Neutrino Physics Centre should offer dedicated computing resources and support to the European neutrino community.
\end{itemize}

The CERN Neutrino Physics Centre would allow Europe to benefit in a visible way from the data collected by the next generation of neutrino experiments, thereby reaping the fruits of the significant investments that are currently being made into the hardware of these experiments. Crucially, if the centre is realised now, with sufficient lead time to establish structures and collaborations before the data arrives, its impact can be expected to peak at the ideal moment, namely at a time when high-luminosity LHC operations will be nearing their end, while a new collider will not be ready yet. 

The CERN Neutrino Physics Centre would build on existing CERN infrastructure (facilities, positions for core staff). What is still required is a long-term commitment to maintaining neutrino physics as an important component of CERN's scientific programme.

\clearpage

\appendix
\section{List of Signatories Including Affiliations}

\makeatletter
\@booleanfalse\@collaboration@present  
\title{}
\let\frontmatter@footnotetext\@gobble   
\let\produce@RRAP\@gobble  
\maketitle


\end{document}